\documentclass[10pt]{article}
\usepackage{amssymb}
\usepackage{amsmath}
\usepackage{indentfirst}
\begin{document}

\title{A new global 1-form in Lyra geometric cosmos model}
\author{Haizhao Zhi$^1$,Mengjiao Shi$^2$,Xinhe Meng*$^3$,Lianzhong Zhang$^4$\\
{\small 1.Department of Physics, Nankai University, Tianjin, China 300071}\\
{\small 2. Chern Institute of Mathematics, Nankai University, Tianjin, China 300071}\\
{\small 3*.Department of Physics, Nankai University, Tianjin 300071, P.R. China,}\\
{\small Kavli Institute of Theoretical Physics China, CAS, Beijing 100190, P.R. China}\\
{\small Email:xhm@nankai.edu.cn (correspondence)}\\
{\small 4.Department of Physics, Nankai University, Tianjin 300071, P.R. China}}
\maketitle

\begin{abstract}Dark energy phenomena has inspired lots of investigations on the cosmological constant problems. In order to understand its origin and properties as well as its impacts on universe's evolutions, there are many approaches to modify the well-known General Relativity, such as the Weyl-Lyra Geometry. In the well studied cosmology model within Lyra geometry, there is a problem that the first law of thermodynamics is violated. To unravel this issue, if we use the effective density and pressure in the Lyra cosmology model to preserve the first law of thermodynamics in the cosmos, the former 1-form $( \beta ,0,0,0)$ cannot give a proper vacuum behavior. In this paper, the auxiliary 1-form is modified to overcome this difficulty. It can be shown that the complex terms in the field equation derived from the regime of Lyra Geometric$ \frac{3}{2}{\phi}^{\mu}{\phi}_{\nu}-\frac{3}{4}{\delta}^{\mu}_{\nu}{\phi}^{\alpha}{\phi}_{\alpha}$with our new 1-form could behave just as the cosmological constant. This work can be regarded as a new exploration on a possible origin of the cosmological constant from a Lyra cosmology model.
\end{abstract}
{\large PACS numbers}\ :98.80.-k 98.80.Qc\\
{\large \ \ Keywords:\ Cosmology;\ Lyra geometry;\ FRW models;\ displacement vector}

\section{Introduction}
Einstein has founded his magnificent theory--the General Theory of  Relativity based on Riemannian geometry, in which the physically gravitational potential is encoded in the geometric metric. In the meantime, there were a lot of framework modifications appeared to have proposed some other possibilities. Weyl [1] has put forward a modified Riemannian geometry with the hope trying to formulate a unified field theory for the gravity and the electromagnetism phenomena, but that attempt failed for the violation of metric preservation. Later, Lyra [2] proposed a new modification of Riemannian geometry with an imposed gauge function which removes the non-integrability by preserving the length of a  parallel transforming vector on the manifold. Then, Sen et al. [3-5] reviewed the geometric theory from Weyl and Lyra. Then, they constructed a scalar tensor theory of gravitation which is in the normal gauge. Soon, Halford [6,7] claimed that the cosmological constant could arise naturally from this 1-form $(\beta,0,0,0)$ in the corresponding cosmology model built from the Lyra's geometry, who has  treated this dual vector as a constant one and shown that the Friedmann equations are almost the same as the ones in the $\Lambda CDM$ model. In the meantime, Herald H. Soleng [8] has purported that this dual vector could be recognized as either the source of the Hoyle's creation field theory [9] or the effective vacuum energy which is similar to the cosmological constant.

On the other hand, many theorists have been studying cosmology by using Lyra geometry framework and got a lot of good results. In the beginning, several authors were unraveling the cosmological constant conundrum with the constant 1-form [10-14]. They all assumed that this 1-form only possess a time component, which is of the form $( \beta ,0,0,0)$. Later, people have realized that this specific 1-form can be time dependent. Singh et al.[15-19], Pradhan et al.[20-31,37] and F. Rahaman et al.[33-36] have studied a lot of Lyra cosmology models based on this assumption considering different Bianchi types of cosmological model or different situations about the viscosity. There are also some work using Lyra manifold as a back ground which couples with other fields[32,50-52], such as Quintessence. People realized that the energy of the Universe should be conserved. Nonetheless, if they use the effective pressure and effective energy to preserve the total energy of the Universe, the vacuum behavior of that model is like a stiff fluid. In our present paper, we propose a possible way to overcome this issue within the framework of Lyra Geometry without invoking other supplementary field.

The present work is arranged as follows: In the second section, we briefly reviewed the mathematical background about Weyl and Lyra geometry, after which in the third section, we discuss the problems of the conventional way to build cosmology model in the Lyra geometry, where we point out clearly that if we want to preserve the first law of thermodynamics, the effective vacuum term will behave as a stiff fluid with an equation of state parameter +1 instead of the -1, which is not the cosmological constant as we have expected. In recent years, several cosmological observations and experiments concerning about the magnitude and the corresponding  red shift of distant type Ia supernova [38-44] showed that there do exist a positive cosmological constant-like term effectively. The existence of this cosmological constant-like term may be the direct reason of why the universe is accelerating. In the following fourth section, we propose a new global 1-form to make sure the first law of thermodynamics satisfied(the energy of the Universe to be conserved). With this method, the effective vacuum in our cosmology model has exactly the same identity as the cosmology constant from the standard Friedmann cosmology model. The
last section is our conclusions and discussions.

\section{Overview of the related mathematical background}
In a pseudo Riemann manifold, the Levi-civita connection $\Gamma^{\alpha}_{\mu\nu}$ is determined uniquely by two requirements: that the length of a vector is not changed under parallel transferring and the connection is torsion-free. Based on Riemannian geometry, Einstein has founded his famous General Theory of Relativity for a new Gravity theory as we know the today. During the past hundred years, many people have tried to modify Riemannian geometry to 
include other forces, like the electric-magnetic phenomena in Kaluza-Klein
theory.
In this present paper we try to construct cosmology model using Lyra geometry via the global 1-form, which is developed after Weyl geometry.

Weyl [1] modified Riemann geometry by introducing Weyl structure on the manifold M. It is a map $F:G \to \varLambda^1(M)$ satisfying $F(e^{\lambda}g)=F(g)-d\lambda$, where $G$ is a conformal structure. It is a an equivalence class of Riemann metrics on M where we say two metrics $g,g'$  are equivalent if and only if $g=e^{\lambda}g'$, $\lambda \in C^{\infty}(M)$. A Weyl manifold is the one with a Weyl structure. We can see that a Riemann metric $g$ and a one-form $\phi_{\mu}dx^{\mu}$ determine a Weyl structure, namely  $F:N \to \varLambda^1(M)$ where $N$ is the equivalence class of $N$ and $F(e^{\lambda})=\phi-d\lambda$.

In a given Weyl manifold $(M,g_{\mu\nu},\phi_{\mu}dx^{\mu})$, there exists a unique torsion-free linear connection compatible with the Weyl structure. The coefficients $\Bar{\Gamma} ^{\alpha}_{\mu\nu}$ of this connection are given by:

\begin{equation}
\Bar{\Gamma} ^{\alpha}_{\mu\nu}=\Gamma^{\alpha}_{\mu\nu}+S^{\alpha}_{\mu\nu}
\end{equation}

\begin{equation}
S^{\alpha}_{\mu\nu}=\frac{1}{2}(\delta^{\alpha}_{\mu}\phi_{\nu}+\delta^{\alpha}_{\nu}\phi_{\mu}-g_{\nu\mu}\phi^{\alpha})
\end{equation}

\begin{equation}
\phi^{\alpha}=g^{\alpha\mu}\phi_{\mu}.
\end{equation}

It is obvious that in common cases under this connection a vector's length is changed after a parallel moving, which is not physically appropriate. As we all know today that the Weyl formulation based on the
Weyl geometry failed to unify the gravity and electromagnetic forces physically, but it is still a very useful subject in mathematics as well as geometry. It is a very intuitive attempt to extend the Riemann geometry.

Motivated by the Weyl geometry, Lyra [2] has defined a displacement vector $PP'$ between two neighboring points $P(x^{\nu})$ and  $P'(x^{\nu}+dx^{\nu})$ by $\delta(PP')=\bar{x}(x^{\nu})dx^{\nu}$, where $\bar{x}$ is called a gauge function. The coordinate system $x^{\nu}$ and a gauge function $\bar{x}$ form a reference system. Without losing generality, we just postulate $\bar{x}=1$.

According to Lyra [2] and Sen [5], we know that in any reference system that the affine connection is determined by $\Gamma^{\alpha}_{\mu\nu}$ and $\phi_{\mu}$ which appears as a consequence of introducing the gauge function $\bar{x}$. The affine connection $^{*}\Gamma^{\alpha}_{\mu\nu}$ are related to the $\Gamma^{\alpha}_{\mu\nu}$ and $\bar{x}$ by:

\begin{equation}
^{*}\Gamma^{\alpha}_{\mu\nu}=\Gamma^{\alpha}_{\mu\nu}+S^{\alpha}_{\mu\nu}
\end{equation}

where $S^{\alpha}_{\mu\nu}$ is defined as above.

The infinitesimal transfer of a vector $\lambda^{\alpha}$ is:

\begin{equation}
\delta\lambda^{\alpha}=-(^{*}\Gamma^{\alpha}_{\mu\nu}-\frac{1}{2}\delta^{\alpha}_{\mu}\phi_{\nu})\lambda^{\mu}\bar{x}dx^{\nu}.
\end{equation}

It can be proved that if the condition $\bar{x}=1$ holds, the length of a vector is not changed under parallel transferring.
Also, the contracted curvature scalar of a Lyra manifold is defined the same as the contracted curvature scalar of Weyl manifold when$\bar{x}=1$. Under the condition $\bar{x}=1$ the curvature scalar of a Lyra manifold is:

\begin{equation}
^{*}R=R+3\phi^{\mu}_{;\mu}+\frac{3}{2}\phi^{\mu}\phi_{\mu}
\end{equation}

where $R$ is the contracted curvature scalar. The volume integral $I$ is given by:

\begin{equation}
I=\int ^{*}R(-g)^{\frac{1}{2}}d^4x
\end{equation}

The field equation could be acquired from the variational principle$\delta(I+J)=0$, where J is volume integral of the Lagrangian density of matter.

Many authors have studied the field equations derived from framework based on the Lyra manifold and try to understand the physics meaning of the 1-form$\phi_{\mu}$, and they all assume the simple form $\phi_{\mu}=(\beta(t),0,0,0)$. We have to say in fact the spatial component of this 1-form can be nonzero naturally.

\section{The Lyra geometry and related gravitational field equations\label{LyraGeom}}

First of all, we rewrite the Einstein field equation without the cosmological constant:
\begin{equation}
G_{\mu\nu}=R_{\mu\nu}-\frac{1}{2}Rg_{\mu\nu}=-\kappa T_{\mu\nu}
\end{equation}
The left hand side (LHS) above is the Einstein Tensor, while the right hand side (RHS) is the corresponding energy-momentum tensor, which is normally assumed as perfect fluid in the standard cosmology. The energy-momentum tensor under our signature is:

\begin{equation}
T^{0}_{0}=\rho,  T^{1}_{1}=T^{2}_{2}=T^{3}_{3}=-p,  T^{\mu}_{\nu}=0   (\mu \ne \nu)
\end{equation}

If we add the cosmological constant $\Lambda$ into the equation above, we shall get:
\begin{equation}
R_{\mu\nu}-\frac{1}{2}Rg_{\mu\nu}+\Lambda g_{\mu\nu}=-\kappa T_{\mu\nu}
\end{equation}
We can write down the equation in space and time components in detail:
\begin{eqnarray}
G_{ii}+\Lambda g_{ii}=-\kappa T_{ii} \    (1\leq i \leq 3)\label{eq:aa1}\\
G_{00}+\Lambda g_{00}=-\kappa T_{00}\label{eq:aa2}
\end{eqnarray}

Secondly, because of our universe is almost statistically isotropic and homogeneous in cosmic large scale, thus the universe space-time symmetry can be depicted by the Friedmann-Roberson-Walker (FRW) metric. The corresponding line element is:

\begin{equation}
ds^2=dt^2-S^2(t)[\frac{dr^2}{1-kr^2}+r^2(d{\theta}^2+sin^2\theta d{\phi}^2)]
\end{equation}
where S(t) is the scale factor and k indicates different signs of curvature (k=0 for flat).
By putting the metric into the Einstein tensor, we could get the explicit form of the component equations accordingly:
\begin{equation}
-G^{1}_{1}=-G^{2}_{2}=-G^{3}_{3}=\frac{k}{S^2}+\frac{{\Dot{S}}^2}{S^2}+\frac{2\Ddot{S}}{S}
\end{equation}

\begin{equation}
-G^{0}_{0}=\frac{3k}{S^2}+\frac{3{\Dot{S}}^2}{S^2}
\end{equation}

Within the regime of Lyra geometry, and following the same standardized method we can get the field equation (or the Einstein-like eq.) in Lyra geometry:

\begin{equation}
G^{\mu}_{\nu}+\frac{3}{2}{\phi}^{\mu}{\phi}_{\nu}-\frac{3}{4}{\delta}^{\mu}_{\nu}{\phi}^{\alpha}{\phi}_{\alpha}=-\kappa\\ T^{\mu}_{\nu}
\end{equation}
In the above equation, the $\phi_\mu$ field is the global 1-form in Lyra geometry.

Under this framework, we could just like everyone else to let $\phi_{\mu}$  be a simple dual vector $\phi_{mu}=( \beta(t),0,0,0)$ which has only nonzero value in the time component, then we plunge this dual vector and the FRW metric into the field equation. We can get the Friedmann equation and the acceleration equation in Lyra geometry cosmology model below:
\begin{equation}
3H^2+\frac{3k}{a^2}-\frac{3}{4}\beta^2=\kappa\rho,
\end{equation}
\begin{equation}
2\dot{H}+3H^2+\frac{k}{a^2}+\frac{3}{4}\beta^2=-\kappa p,
\end{equation}
where $H=\frac{\dot{S}}{S}$ is the Hubble's parameter.
From these two equations, we could derive the continuity equation, which reads:
\begin{equation}
\dot{\rho}+\frac{3}{2\kappa}\beta\dot{\beta}+3H(\rho+p+\frac{3}{2\kappa}\beta^2)=0.
\end{equation}
Then we moved the $\frac{3}{4}\beta^2$ term in the Friedmann equation and the acceleration equation to the RHS of their equations. We translate the combination of the right hand side as effective density and effective pressure separately. These two field equations then will be presented as:
\begin{equation}
3H^2+\frac{3k}{a^2}=\kappa\rho_{ef},
\end{equation}
\begin{equation}
2\dot{H}+3H^2+\frac{k}{a^2}=-\kappa p_{ef}.
\end{equation}
At this time, we use the effective density and effective pressure to substitute the counterparts in the continuity equation, then the continuity equation will be:
\begin{equation}
\dot{\rho_{ef}}+3H(\rho_{ef}+p_{ef})=0.
\end{equation}
We define the effective density and effective pressure as below:
\begin{equation}
\rho_{ef}=\rho+\frac{3\beta^2}{4\kappa},
\end{equation}
\begin{equation}
p_{ef}=p+\frac{3\beta^2}{4\kappa}.
\end{equation}
This procedure is necessary, because the former continuity equation obviously violates the first law of thermodynamics. We consider the cosmos possess a volume with a co-moving radius 1, then the universe has a physical radius S if we assume the Universe is flat. We can represent the energy of the Universe as $E=\frac{4\pi}{3}S^3\rho.$
Because the scale factor and the density are functions of time, we could derive that:
\begin{equation}
dE=4\pi S^2\rho dS+\frac{4\pi}{3}S^3d\rho
\end{equation}
The differential change of volume is: $dV=4\pi S^2dS$.

We could just focus on a recent cosmology process, which might be the cosmic expansion period from the redshift $z\approx 0.5$ to today's universe.  Also, the energy of the universe is always conserved during the expansion. Then, we could suppose this specific process is adiabatic  which preserves the entropy. So, from the first law of thermodynamics that $dE+pdV=0$. We then put everything into the first law of thermodynamics equation and the final result above is thus:

\begin{equation}
\dot{\rho}+3H(\rho+p)=0
\end{equation}

So, only the effective energy and pressure could render the appropriate fluid continuous equation.

Then, it is appropriate to consider the state equation parameter $w$ as the quotient of the effective pressure dividing the effective density:

\begin{equation}
w=\frac{p_{ef}}{\rho_{ef}}=-1-\frac{2\dot{H}}{3H^2}.
\end{equation}

So, first of all, we should consider the meaning of the cosmological constant. Because the covariant derivation of the metric tensor is zero, the presence of the cosmological constant term in the energy-momentum tensor does not violate the first law of thermodynamics. The existence of the cosmological term manifests that even there is no matter components in the cosmos, the cosmological term will also have its own energy density with the $w=-1$ as the equation of state parameter value. This cosmological constant has been identified as the vacuum energy, so-called the Zero Point Energy[45-47].

But if we look at the vacuum of the cosmology described by Lyra geometry, picking the displacement vector as usual, and make the density and pressure both zero, then the equation of state parameter (EoS) w=+1, which is the stiff fluid and is unacceptable for the dark energy dominated universe today. It is the existence of the cosmological constant with  EoS w=-1 that makes the universe accelerating instead of  the cosmic stiff fluid component. With various astrophysics observations such as the the cosmic microwave background anisotropy, large scale structure survey, gravitational lensing and the recent
 supernova experiments and observations, like the Union2.1 compilation [48,49] shown,  have revealed that the flat universe is with the best fitted constant equation of state parameter $w=-0.997^{+0.050}_{-0.054}$ and the universe with curvature (not necessarily with flat geometry) has got  the equation of state parameter $w=-1.035^{+0.055}_{-0.059}$. In the meantime, the vacuum energy has the energy-momentum tensor just as the cosmological constant term described, whose EoS is exactly $w=-1$. To reconcile this observational data, we have introduced a new global 1-form which could attain a good result.

\section{New displacement vector cosmology model}
Instead of the simply chosen vector ($\beta$,0,0,0) which only possess a time-component, we extend this 4-vector to involve nonzero spatial components. We just postulate that the displacement vector has got the form:
\begin{equation}
{\phi}^{\mu}=(\beta ,\alpha_1,\alpha_2, \alpha_3)
\end{equation}
In which $\alpha _i$ and $\beta$ are functions of time and space. This displacement vector is the dual of the global 1-form.
With the introduced FRW metric, immediately, we can derive that:
\begin{equation}
\phi^0=\phi_0=\beta
\end{equation}

\begin{equation}
\phi_1=g_{11}\phi^1=\frac{-S^2(t)}{1-kr^2}\alpha_1
\end{equation}

\begin{equation}
\phi_1 \phi ^1=-{\alpha_1}^2\frac{S^2(t)}{1-kr^2}
\end{equation}

Following the same procedure for getting the equation (31) above, we can easily get:
\begin{equation}
\phi_2 \phi^2=-{\alpha_2}^2{S^2(t)}r^2
\end{equation}
\begin{equation}
\phi_3 \phi^3=-{\alpha_3}^2S^2(t)r^2sin^2\theta
\end{equation}

We just put (29),(31),(32),(33) into the field equation (16) derived from the Lyra geometry. Here come out four equations. Each equation corresponds to a specific space-time component:

\begin{equation}
G^{1}_{1}+\frac{3}{2}\phi_1 \phi ^1-\frac{3}{4}\phi^{\alpha}\phi_{\alpha}=-\kappa T^{1}_{1}
\end{equation}

\begin{equation}
G^{2}_{2}+\frac{3}{2}\phi_2 \phi ^2-\frac{3}{4}\phi^{\alpha}\phi_{\alpha}=-\kappa T^{2}_{2}
\end{equation}

\begin{equation}
G^{3}_{3}+\frac{3}{2}\phi_3 \phi ^3-\frac{3}{4}\phi^{\alpha}\phi_{\alpha}=-\kappa T^{3}_{3}
\end{equation}

\begin{equation}
G^{0}_{0}+\frac{3}{2}\phi_0 \phi ^0-\frac{3}{4}\phi^{\alpha}\phi_{\alpha}=-\kappa T^{0}_{0}
\end{equation}

We expect the new 1-form could reflect the characteristics of the cosmological constant term, we get:
\begin{equation}
\phi_1 \phi ^1=\phi_2 \phi ^2=\phi_3 \phi ^3
\end{equation}

Then, the first three equations (34),(35) and (36) are the same, which is what we have expected. We focus on Eqs.(34),(37). Then we get the field equations in the explicit form:
\begin{equation}
\frac{k}{S^2}+\frac{{\Dot{S}}^2}{S^2}+\frac{2\Ddot{S}}{S}+\frac{3}{4}\phi_1 \phi ^1+\frac{3}{4}\beta^2=-\kappa p
\end{equation}
and
\begin{equation}
\frac{3k}{S^2}+\frac{3{\Dot{S}}^2}{S^2}+\frac{9}{4}\phi_1 \phi ^1-\frac{3}{4}\beta^2=\kappa \rho
\end{equation}

Compared with the relativistic cosmology equations with the cosmological constant term:
\begin{equation}
\frac{k}{S^2}+\frac{{\Dot{S}}^2}{S^2}+\frac{2\Ddot{S}}{S}-\Lambda=-\kappa p
\end{equation}

\begin{equation}
\frac{3k}{S^2}+\frac{3{\Dot{S}}^2}{S^2}-\Lambda=\kappa \rho
\end{equation}

So if the condition $\frac{9}{4}\phi_1 \phi ^1-\frac{3}{4}\beta^2=\frac{3}{4}\phi_1 \phi ^1+\frac{3}{4}\beta^2$ meets, the equations (39),(40) and the equations(41),(42) are intrinsically identical when concerning on the cosmological constant term. Now we have that $\frac{9}{4}\phi_1 \phi ^1-\frac{3}{4}\beta^2=\frac{3}{4}\phi_1 \phi ^1+\frac{3}{4}\beta^2$, which means$\phi_1\phi ^1=\beta^2$.
We can simplify the equations we have just got above. Then we can rewrite down the new equations corresponding to equations (39) and (40) respectively.
\begin{equation}
\frac{k}{S^2}+\frac{{\Dot{S}}^2}{S^2}+\frac{2\ddot{S}}{S}+\frac{3}{2}\beta^2=-\kappa p
\end{equation}
\begin{equation}
\frac{3k}{S^2}+\frac{3{\Dot{S}}^2}{S^2}+\frac{3}{2}\beta^2=\kappa \rho
\end{equation}

Just as the method has been used above, we can derive the new effective pressure $p_{ef}^{*}$ and the effective density $\rho_{ef}^{*}$ as now:
\begin{equation}
p_{ef}^{*}=p+\frac{3\beta^2}{2\kappa},
\end{equation}
\begin{equation}
\rho_{ef}^{*}=\rho-\frac{3\beta^2}{2\kappa}.
\end{equation}
These two new quantities satisfy the continuity equation
\begin{equation}
\dot{\rho_{ef}}+3H(\rho_{ef}^{*}+p_{ef}^{*})=0,
\end{equation}
which obviously implies the first law of thermodynamics still holds.

At last we consider the equation of state parameter $w=\frac{p_{ef}^{*}}{\rho_{ef}^{*}}=\frac{p+\frac{3\beta^2}{2\kappa}}{\rho-\frac{3\beta^2}{2\kappa}}$.
Under the vacuum condition:$\rho =p=0$, it is obvious that$w=-1$ which is exactly the same as the cosmological constant EoS.

So we should have $\frac{3}{2}\beta^2=-\Lambda$ and the $\Lambda$ is regarded as the cosmological constant. Now,
we can write out the displacement vector explicitly, from the condition $\phi_1 \phi ^1=\phi_2 \phi ^2=\phi_3 \phi ^3=\beta^2=-\frac{2}{3}\Lambda$.

\begin{equation}
\beta=\sqrt{\frac{2}{3}\Lambda}i
\end{equation}

\begin{equation}
\alpha_1=\frac{\sqrt{\frac{2}{3}\Lambda(1-kr^2)}}{S(t)}
\end{equation}

\begin{equation}
\alpha_2=\frac{\sqrt{\frac{2}{3}\Lambda}}{S(t)r}
\end{equation}

\begin{equation}
\alpha_3=\frac{\sqrt{\frac{2}{3}\Lambda}}{S(t)rsin\theta}
\end{equation}

This 4-vector here is analogous to the result of a wick rotation which the inner product of a 4-vector is transferred from the Minkowski metric convention to the 4-dimensional Euclidean metric convention, if one permits that the coordinate t to take on imaginary values. Like the 4-vector in wick rotation, the time component of our displacement vector here is imaginary. Thus, the new Lyra displacement vector consolidated is:
\begin{equation}
\phi^{\mu}=\frac{\sqrt{\frac{2}{3}\Lambda}}{S(t)}(S(t)i,\sqrt{1-kr^2},\frac{1}{r},\frac{1}{rsin\theta})
\end{equation}
The cosmological constant from this new model is equal to $\Lambda=-\frac{9}{4}\phi_1 \phi ^1+\frac{3}{4}\beta^2=-\frac{3}{4}\phi_1 \phi ^1-\frac{3}{4}\beta^2=-\frac{3}{2}\beta^2$. The cosmological constant here is real and positive.

\subsection{The vacuum Friedman equation}
As a result for application to currently accelerating expansion universe, which can be described by a relatively mathematically simple cosmology model, we can get the vacuum Friedman equation from Eq.(44) directly, or when the cosmological constant-like dark energy dominated universe phase
\begin{equation}
H^2+\frac{k}{S^2}+\frac{\beta^2}{2}=0.
\end{equation}
Where H is Hubble parameter. So,  Eq.(53) could be used to describe the de Sitter universe evolution phase. By neglecting the influence of matter (mainly cold dark matter) and radiation, the vacuum energy is the dominated fraction in the universe media for cosmic evolution, which is also adaptable to the inflation stage of the evolution for the very early universe.

The solutions of Eq.(53) are thus respectively for various geometries
\begin{itemize}
\item $k=0,\ S(t)=mexp(\frac{\sqrt{\Lambda}}{3}t)$ (with m as an integral constant)
\item $k=+1,\ S(t)=\frac{3}{\sqrt{\Lambda}}cosh(\frac{\sqrt{\Lambda}}{6}t)$
\item $k=-1,\ S(t)=\frac{3}{\sqrt{\Lambda}}cosh(\frac{\sqrt{\Lambda}}{6}t)$
\end{itemize}
The value of $\Lambda$ above as a kind vital energy in universe is  detected now as in [49]: $\Lambda \sim 10^{-83}Gev^2$.

\section{Discussions and conclusions}
In the present work, extending the 1-form in the framework of Lyra geometry could solve the issue in the original Lyra cosmology model which only concerned about the displacement vector with only the time component $(\beta,0,0,0)$. The problem is that if we want to preserve the conservation of the first law of thermodynamics, we have to introduce the effective pressure and the effective density. But if we concentrate on the so-called vacuum energy dominated situation, the equation of state parameter $w=+1$ which is the stiff fluid, instead the required cosmological constant-like term. This is then invalid. We have focused on this issue and solved it with the expansion of the global 1-form. Our new auxiliary global 1-form has its effective terms$ \frac{3}{2}{\phi}^{\mu}{\phi}_{\nu}-\frac{3}{4}{\delta}^{\mu}_{\nu}{\phi}^{\alpha}{\phi}_{\alpha}$ equal to the cosmological constant term exactly. At the same time, the first law of thermodynamics still holds.

Using the vector field to describe an effective vacuum for the dark energy physics, like the ether field discussed widely as in [53,54] and references therein, has got a long and interesting history. We will leave the related work for another exploration elsewhere later. Dark energy mystery, together with the dark matter puzzles, has been inspiring us for more probes of astrophysics, both observationally and theoretically with the hope that efforts and endeavor can bring us new ideas for fundamental physics at least.  
\begin{center}
\textbf{Acknowledgments}
\\
\end{center}
We would like to express our gratitude to Sanjoy Kumar Routh, a graduate student presently at the 'S N Bose National Centre for Basic Sciences' from India, for his pointing out a minor mistake to us in our calculations. This work is partly supported by Natural Science Foundation of China under Grant Nos.11075078 and 10675062, and by the project of knowledge Innovation Program (PKIP) of Chinese Academy of Sciences (CAS) under the grant No. KJCX2.YW.W10 through the KITPC astrophysics and cosmology programm where we have initiated this present work.
\\

\textbf{Reference}

[1] H. Weyl, Sber. Preuss. Akad. Wiss. (Berlin) 465 (1918).

[2] G. Lyra, Math. Z. 54, 52 (1951).

[3] D. K. Sen, Z. Phys. 149, 311 (1957).

[4] D. K. Sen and K. A. Dunn, J. Math. Phys. 12, 578 (1971).

[5] D. K. Sen and J.R.Vanstone, J.Math.Phys. Vol.13, No.7,July 1972

[6] W. D. Halford, Aust. J. Phys. 23, 863 (1970).

[7] W. D. Halford, J. Math. Phys. 13, 1699 (1972).

[8] H. H. Soleng, Gen. Relativ. Grav. 19, 1213 (1987).

[9] Hoyle, F. (1948). Mon. Not. Roy. Astr. Soc., 108,372.

[10] K. S. Bhamra, Austr. J. Phys. 27, 541 (1974).

[11] T. M. Karade and S. M. Borikar, Gen. Rel. Gravit. 9, 431 (1978).

[12] S. B. Kalyanshetti and B. B. Waghmode, Gen. Rel. Gravit. 14, 823 (1982).

[13] D. R. K. Reddy and P. Innaiah, Astrophys. Space Sci. 123, 49 (1986).

[14] D. R. K. Reddy and R. Venkateswarlu, Astrophys. Space Sci. 136, 183 (1987).

[15] T. Singh and G. P. Singh, J. Math. Phys. 32, 2456 (1991a).

[16] T. Singh and G. P. Singh, Il. Nuovo Cimento B 106, 617 (1991b).

[17] T. Singh and G. P. Singh, Int. J. Theor. Phys. 31, 1433 (1992).

[18] T. Singh and G. P. Singh, Fortschr. Phys. 41, 737 (1993).

[19] G. P. Singh and K. Desikan, Pramana-journal of physics, 49, 205 (1997).

[20] A. Pradhan, V. K. Yadav, and I. Chakrabarty, Int. J. Mod. Phys. D 10, 339 (2001).

[21] A. Pradhan and A. K. Vishwakarma, SUJST XII Sec. B, 42 (2000).

[22] A. Pradhan and I. Aotemshi, Int. J. Mod. Phys. D 9, 1419 (2002).

[23] A. Pradhan and A. K. Vishwakarma, Int. J. Mod. Phys. D 8, 1195 (2002).

[24] A. Pradhan and A. K. Vishwakarma, J. Geom. Phys. 49, 332 (2004).

[25] A. Pradhan 2009 Commun. Theor. Phys. 51 378 doi:10.1088/0253-6102/51/2/38

[26] A. Pradhan, S.S.Kumhar, Astrophys Space Sci. 54, 137-146 (2009).

[27] A. Pradhan, P. Mathur, Fizika B. 18, 243-264 (2009).

[28] A. Pradhan, P. Yadav, Int. J. Mathematical Sci. (2009), DOI: 10.1155/2009/471938.

[29] A. Pradhan, J. Math. Phys. 50, 022501-022513.23 (2009).

[30] A. Pradhan, H. Amirhashehi, H. Zanuddin, IJTP 50, 56-69 (2011).

[31] A. Pradhan, A.K.Singh, IJTP 50, 916-933 (2011).

[32] Raj Bali et al 2010 Commun. Theor. Phys. 54 197 doi:10.1088/0253-6102/54/1/36

[33] F. Rahaman, Int. J. Mod. Phys. D 10, 579 (2001).

[34] F. Rahaman, S. Chakraborty, and M. Kalam, Int. J. Mod. Phys. D 10, 735 (2001).

[35] F. Rahaman, S. Das, N. Begum, and M. Hossain, Pramana 61, 153 (2003).

[36] F. Rahaman, Il Nuovo Cimento B 118, 99 (2003).

[37] S. Agarwal, R.K. Pandey, A. Pradhan, IJTP 50, 296-307 (2011).

[38] P. M. Garnavich et al., Astrophys. J. 493, L53 (1998a).

[39] P. M. Garnavich et al., Astrophys. J. 509, 74 (1998b).

[40] S. Perlmutter et al., Astrophys. J. 483, 565 (1997).

[41] S. Perlmutter et al., Nature 391, 51 (1998).

[42] S. Perlmutter et al., Astrophys. J. 517, 565 (1999).

[43] A. G. Riess et al., Astron. J. 116, 1009 (1998).

[44] B. P. Schmidt et al., Astrophys. J. 507, 46 (1998).

[45] S.Weinberg, Rev. Mod. Phys.61, 1-23 (1989)

[46] Meng Xinhe, and Peng Wang, Classical and Quantum Gravity 22.1 (2005).

[47] Saibal Ray, Utpal Mukhopadhyay,Xin-He Meng, Cosmology, No. 2 (50), pp. 142¨C150, Vol. 13 (2007).

[48] R. Amanullah et al, Astrophys. J. 716 (2010) 712.

[49] Jorge L. Cervantes-Cota, George Smoot,  arXiv:1107.1789.

[50] V. K. Shchigolev, arXiv:1207.5476.

[51] V. K. Shchigolev, Mod. Phys. Lett. A, 27, 1250164 (2012).

[52] V. K. Shchigolev, E. A. Semenova , arXiv:1203.0917.

[53] X.H.Meng and X. L. Du, Phys.Lett.B 710,493(2012);

[54] Meng Xin-He and Du Xiao-Long 2012 Commun. Theor. Phys. 57 227 doi:10.1088/0253-6102/57/2/12


\end{document}